\documentstyle[aps,prb,epsf]{revtex}
\def\re{\Re{\rm e}}

\def\tf{\theta}
\def\sgn{\mathop{\rm sgn}\nolimits}
\begin{document}

\twocolumn[\hsize\textwidth\columnwidth\hsize\csname%
  @twocolumnfalse\endcsname 

\title{Coulomb interactions and delocalization in quantum Hall
  constrictions.}
  
\author{Leonid P. Pryadko$^*$, Efrat Shimshoni$^\dagger$, Assa
  Auerbach$^\ddagger$} 
\address{$^*$Institute for Advanced Study, Princeton, NJ 08540.}
\address{$^\dagger$ Department of Mathematics-Physics, 
  Oranim--Haifa University, Tivon 36006, Israel.}
\address{$^\ddagger$ Department of Physics, The Technion, 
  Haifa 32000, Israel.}
\date\today 
\maketitle
\begin{abstract}
  We study a geometry-dependent effect of long-range Coulomb
  interactions on quantum Hall (QH) tunneling junctions.  In an {\sf
    X}-shaped geometry, duality relates junctions with opening angles
  $\alpha$ and $(\pi-\alpha)$.  We prove that duality between weak
  tunneling and weak backscattering survives in the presence of long
  range interactions, and that their effects are precisely cancelled
  in the self dual geometry $\alpha=\pi/2$.  Tunneling exponents as a
  function of $\alpha$, the interaction strength $\chi$ and the
  filling fraction $\nu$ are calculated.  We find that Coulomb
  interaction induces localization in narrow channels (large
  $\alpha$), and delocalization for sharply pinched constrictions
  (small $\alpha$).  Consequently, an insulator to metal transition
  happens at an angle $\alpha_c(\chi,\nu)\le \pi/2$.  We discuss the
  implications of our results for tunneling experiments in
  QH-constriction and cleaved-edge geometries.
\end{abstract}

\pacs{}]
\narrowtext
\section{Introduction}

The idea of current-carrying edge states\cite{Halperin-82} is one of
the major paradigms in the theory of the quantum Hall (QH) effect.
For simple filling fractions $\nu=(2m+1)^{-1}$, Wen has
shown\cite{Wen-90,Wen-91,Wen-91A,Wen-91B,Wen-92rev} that edge modes
can be represented as one-component chiral Luttinger liquids, with the
universal coupling determined by $\nu$.  Within this simple model,
controlled calculations are possible.  This lead to many beautiful
results, including the universal inter-edge tunneling
exponent\cite{Wen-91B,Kane-Fisher-Tunnel}, exact expressions for
tunneling conductance, the non-linear tunneling $I$--$V$
curve\cite{Weiss-Exact,Fendley-95B}, and tunneling
noise\cite{Kane-94A,Fendley-95C,QHpers-book}.
 
Experimentally, however, there are more dimensions to this problem.
The results of the first pinch-off tunneling
experiment\cite{Milliken-96}, where the scaling appeared to be in
agreement with theory\cite{Wen-91B,Kane-Fisher-Tunnel}, have only
recently received a partial confirmation\cite{Maasilta-97,Turley-98}.
Furthermore, in Ref.~\CITE{Ando-98} no scaling was observed at all,
and in Ref.~\CITE{Glattli-99} the measured
tunneling exponent was off by a factor of two. 
Such discrepancies were attributed in part to edge
reconstruction in samples with ``soft'' confinement\cite{softedge}.
However, the tunneling measurements in cleaved-edge
samples\cite{Chang-96,Grayson-98}, where the confining potential is
expected to be sharp, yield tunneling exponents shifted off the
predicted values even at the magic filling fractions $\nu=1$, $1/3$.

\begin{figure}[htbp]
  \begin{center}
    \leavevmode \epsfxsize0.75\columnwidth
    \epsfbox{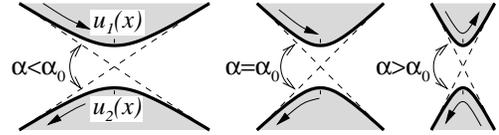}\vskip0.2pc
    \caption{
      Shading denotes quantum Hall regions bounded by
      conterpropagating edge modes $u_1,u_2$. In a saddle point
      geometry, Coulomb interactions suppress the up-down tunneling
      for large opening angles $\alpha$, and enhance it for small
      $\alpha$.  This effect cancels {\em exactly\/} in the self-dual
      geometry, $\alpha=\pi/2$.}
    \label{fig:angles}
  \end{center}
\end{figure}

Previously, much effort\cite{mechanisms} was dedicated to identify
mechanisms leading to (non-universal) corrections to tunneling
exponents.  In particular, the effect of the long-range Coulomb
interaction was analyzed\cite{Zuelicke-96,Moon-96,Oreg-96,Imura-97} in
the geometry of two counterpropagating parallel edges ($\alpha\to0$ in
Fig.~\ref{fig:angles}).  In exact analogy with its effect in
one-dimensional electron gas\cite{Emery-1DEG}, repulsive Coulomb
interaction renormalizes the Luttinger liquid coupling parameter.
Thus, a weak impurity-associated inter-edge tunneling becomes a
relevant perturbation, so that the current flow (from top to bottom in
Fig.\ \ref{fig:angles}) is {\em enhanced} at low temperature $T$ and
applied voltage $V$. However, the same interaction {\em
  suppresses}\cite{Imura-97} the tunneling in the dual configuration,
of two semi-infinite non-chiral Luttinger liquids connected by a
tunneling point ($\alpha\to\pi$ in Fig.~\ref{fig:angles}), and the
system is pushed towards the insulating regime.  This indicates that
even the {\em sign\/} of the Coulomb interaction effect on the
tunneling exponent is not the same in different geometries.

The purpose of this work is to analyze in detail the Coulomb
interaction effect on the properties of QH tunneling junctions in
different geometries.  First, we demonstrate that the well-known
duality relating weak tunneling and weak backscattering remains exact
in the presence of long-range interactions.  Then, we focus on
scale-invariant {\sf X}-shaped constrictions, and calculate the
renormalized Luttinger coupling constant 
$g_\star^2$ (which, in particular, determines the power law
dependence of the conductance on $T$ and $V$)
as a function of the opening angle
$\alpha$ (Fig.~\ref{fig:angles}).  We show that the 
unscreened Coulomb interaction drives a zero temperature
delocalization transition as a function of $\alpha$ in both integer
and fractional QH constrictions. In the integer case the
transition occurs precisely at the self-dual value $\alpha_c=\pi/2$,
independent of the interaction strength. At the fractions 
$\nu=(2m+1)^{-1}$, the critical angle $\alpha_c$ is non-universal, 
but its value is always smaller than $\pi/2$.  
We also analyze the effect of Coulomb
interactions in the geometry of cleaved-edge tunneling experiments.

The paper is organized as follows.  In
Sec.~\ref{sec:effective-tunneling} we introduce the tunneling action
which accounts for the long-range interactions.  A general proof of
the duality between weak tunneling and weak backscattering is given in
Sec.~\ref{sec:duality}.  In Sec.~\ref{sec:self-similar}, we present
our results for the renormalized Luttinger coupling $g_\star^2$ in different
geometries, and in Sec.~\ref{sec:discussion} we discuss the
implications on tunneling experiments.  Related analytic results
are collected in Appendices: in App.~\ref{sec:appendix-pi}, the case of
$\alpha=\pi$ is solved; in App.~\ref{sec:wiener-hopf}, the Wiener-Hopf 
technique is used to directly solve the self-dual case $\alpha=\pi/2$, 
and evaluate the lowest order correction for $|\cos\alpha|\ll1$. 

\section{The effective tunneling action}
\label{sec:effective-tunneling}

Gapless edge excitations $u\equiv u(x,\tau)$ for Laughlin's QH states
with filling fractions $\nu\!=\!(2m+1)^{-1}$ can be
described\cite{Wen-91A,Wen-91B,Wen-92rev} by the imaginary-time
quadratic action
\begin{equation}
  \label{eq:edge-action}
  {\cal S}_0=\frac{1}{4\pi}\int_0^\beta \!d\tau\int dx\,{\partial_x
    u\,(i\partial_\tau{u}+ v \,\partial_x u)},
\end{equation}
where $x$ is the coordinate along the edge, and $v\equiv v(x)$ is the
edge wave velocity. The field $u$ is related to the linear charge
density at the edge, $\rho=\sqrt\nu\,\partial_x u/(2\pi)$ (note the
unconventional normalization).

Formally, gauge invariance requires that the field $u(x,\tau)$ be
treated as a compact boson of radius $R=\sqrt\nu$, {\em i.e.}, the
values $u$ and $u+2\pi\sqrt\nu$ must be identified.  This, however, is
{\em not\/} achieved within the usual path integral
formalism\cite{Oreg-95} in a finite geometry if we assume the field
$u(x,\tau)$ continuous everywhere along the circumference.  Indeed,
the equal-time commutation relationship
$$
[u(x),\,u(x')]=i\pi\sgn (x-x')
$$
on the edge of length $L$ implies that the fields $u_0\equiv
u(0,\tau)$ and $u_L\equiv u(L,\tau)$ are canonically conjugated, which
contradicts the continuity of the field along the circle.  The
difference $u_L-u_0$ (proportional to the topological charge
associated with the zero mode) is also proportional to the total
charge $Q=\sqrt\nu\,(u_L-u_0)/(2\pi)$ accumulated at the edge; only in
the absence of tunneling into the edge this charge is a dynamically
conserved quantized quantity.  The correct zero-mode quantization
spectrum can be obtained if we consider the variables $u_0$ and $u_L$
as independent, and write the bare edge action~(\ref{eq:edge-action})
more explicitly as\cite{Pryadko-98}
\begin{eqnarray}
  {\cal S}_0& =& \frac{1}{4\pi}\int_{0\strut}^\beta d\tau\int_0^L
  dx\,{\partial_x u\,(i\partial_\tau{u}+ v
    \,\partial_x u)}\nonumber \\
  & & +\,{1\over8\pi}\int_0^{\beta\strut} d\tau \,(u_L-u_0)\,
  i\partial_\tau (u_L+u_0).
  \label{eq:finite-size-edge-action}
\end{eqnarray}
The boundary term in the second line is added to fix the canonical
quantization of the zero mode, and to decouple it from the edge modes
with finite momenta.

Since the charge density $\rho$ is expressed linearly in 
terms of the field $u$, the action remains
quadratic\cite{Wen-92rev,Moon-96,Hangmo-96} even in the presence
of non-local Coulomb interaction
\begin{equation}
  \label{eq:coulomb-action}
  {\cal S}_1= {\nu\,e^2\over  8\pi^2\varepsilon}\int_0^\beta
  \!d\tau\int  dx\,dy \, 
  u'(x)\,V\left(\left|{\bf r}_x-{\bf r}_y\right|\right)\,u'(y),
\end{equation}
where ${\bf r}_x$ is the actual position of the point $x$ as measured
along the edge, and $\varepsilon$ is the dielectric constant of the
material.  The problem is non-trivial because now both the distance
$x$ measured along the edge, and the geometrical distance $\left|{\bf
    r}_x-{\bf r}_y\right|$ are important.

The inter-edge tunneling is introduced by the
non-linear term
\begin{equation}
  \label{eq:tunn-action}
  {\cal S}_{\rm t}=\int_0^\beta \! d\tau \,
  \re\,\lambda\,e^{i g \varphi}, \quad \varphi\equiv u(x_1)-u(x_2);
\end{equation}
here $g\!=\!\sqrt\nu$ for the qua\-si\-par\-ti\-cles' tunneling
between the points $x_1$ and $x_2$ through the QH liquid with the
fill\-ing fraction $\nu$, or $g\!\rightarrow \!\tilde g=1/\sqrt{\nu}$
for tunneling of electrons through the insulating region.  The
tunneling amplitude $\lambda$ is set by the
details\cite{Jain-88} of the self-consistent potential near the
tunneling point and considered as a phenomenological parameter.

The non-linear tunneling action~(\ref{eq:tunn-action}) depends on the
values of the field $u(x,\tau)$ in the points $x_1$, $x_2$; the values
of this field in all other points can be integrated out.  Leaving the
argument $\varphi$ of the tunneling term as the only independent
variable, we can write the most general form of the effective action
\begin{equation}
  \label{eq:effective-model}
  S={T\over4\pi}\sum_{n} %
  |\omega_n|
  \,{\cal K}(\omega_n)\,|\varphi_n|^2 
  +\int_0^\beta \! d\tau \,
  \re\,\lambda\,e^{i g \varphi(\tau)},
\end{equation}
where the harmonics $\varphi_n\equiv \int_0^\beta d\tau
\varphi(\tau)\,\exp(-i\omega_n\tau)$ and $\bar\varphi_n\equiv
\varphi_{-n}$ are evaluated at the Matsubara frequencies
$\omega_n=2\pi\,n T$.  This effective tunneling model is fully
characterized by the frequency-dependent coupling ${\cal
  K}(\omega_n)$, which contains all relevant information about the
form of the interaction potential $V(r)$ and the geometry of the
system.  Formally, its functional form is defined by the
correlator\cite{Pryadko-98}
\begin{equation}
  \label{eq:thermal-average}
  {\cal K}^{-1}(\omega_n) ={|\omega_n|\over2\pi} 
  \bigl\langle\left|\varphi_n\right|^2\bigr\rangle_{\lambda=0}.
\end{equation}

If the coupling ${\cal K}(\omega)$ is independent of the frequency,
the effective action~(\ref{eq:effective-model}) can be visualized as
describing an overdamped particle in a periodic (cosine) potential
with Ohmic dissipation $\kappa={\cal K}/g^2$; the transport properties
for this problem are known exactly\cite{Weiss-Exact,Fendley-95B}.  In
 general, however, the exact solution is not available, and we
have to rely on the frequency-shell perturbative renormalization group
(RG).  The main idea is that the non-linear term is irrelevant for
large-frequency modes $\varphi(\omega)$, as long as $|\omega|\gg
\lambda$.  When such modes are integrated out, the tunneling constant
for the remaining slow modes is reduced,
\begin{equation}
  \label{eq:eff-tunn}
  \lambda(\Lambda) =\lambda(\Lambda_0)\left\langle 
    e^{ig\varphi}
  \right\rangle_{\Lambda<\omega<\Lambda_0}, 
\end{equation}
or, equivalently, 
$$
-\ln{\lambda(\Lambda)\over \lambda(\Lambda_0)}=
{g^2}\int_{\Lambda_0}^\Lambda {d\omega\over2\pi} \left\langle
  |\varphi(\omega)|^2\right\rangle_{\lambda=0}
={g^2}\int_{\Lambda_0}^\Lambda {d\omega\over\omega{\cal K}(\omega)},
$$
where we used the definition~(\ref{eq:thermal-average}).  After the
frequencies are rescaled to restore the original upper cutoff, we
arrive at the usual RG equation
\begin{equation}
{d\ln\lambda\over d\ln\Lambda} =1-g^2\,{\cal K}^{-1}(\Lambda) \equiv
1-g_\star^2(\Lambda).
\label{eq:define-gstar} 
\end{equation}
The renormalization stops at a lower cutoff scale determined either by
the temperature or the applied voltage.  Most importantly, for
$g_\star^2 >1$, the tunneling amplitude flows to weak coupling as the
temperature is lowered, so that the channel along the tunneling
current becomes more insulating; for $g_\star^2 <1$ it flows to strong
coupling.

It should be pointed out that in the case where ${\cal K}(\omega)$ is
{\em frequency-independent\/}, the parameter $g_\star^2$ [defined in
Eq.~(\ref{eq:define-gstar})] is a constant, and the effective
Euclidean action describing the system can be recast in the simpler
form
\begin{equation}
  S={T\over4\pi}\sum_{n} |\omega_n|\,|\varphi_n|^2 
  +\int_0^\beta \! d\tau \,
  \re\,\lambda\,e^{i g_\star \varphi(\tau)}. 
  \label{eq:effective-action-disc}
\end{equation} 
Such is indeed the case (for sufficiently small $\omega$) for the
scale-invariant models considered in detail in
Sec.~\ref{sec:self-similar}.  In this situation, the RG equation leads
to the standard result\cite{Kane-Fisher-Tunnel,QHpers-book}
\begin{equation}
  \lambda_{\rm eff}\sim \max(T,\,V)^{g_\star^2-1}, 
  \label{eq:lameff}
\end{equation}
which can be also obtained by expanding the exact
solution\cite{Weiss-Exact,Fendley-95B}.

\section{Duality between weak tunneling and weak bacscattering}
\label{sec:duality}

The partition function corresponding to the effective
action~(\ref{eq:effective-model}) [which also describes an overdamped
  particle in a non-Ohmic dissipative environment,
  $\kappa(\omega)={\cal K}(\omega)/g^2$] can be also
rewritten\cite{Schmid-83,Guinea-85} in terms of the dual variable
$\Delta\theta$ with the identical action, up to a replacement ${\cal
  K}(\omega_n)\to 1/{\cal K}(\omega_n)$, $g\to 1/g$, and the modified
tunneling coefficient $\lambda\to \tilde\lambda$ (which has the
meaning of fugacity for the instanton of the original field
$\varphi$).  In terms of edge modes, this
duality\cite{Fendley-95B,Weiss-Exact} represents a freedom to describe
the same junction in terms of {\em weak\/} tunneling or {\em strong\/}
backscattering, and vice versa.  The main advantage of the duality is
the ability to substitute a problem at {\em strong\/} tunneling with
its dual, which can be then accessed perturbatively.

This argument relies heavily on the properties of the effective
model~(\ref{eq:effective-model}), which, in principle, may or may not
remain equivalent to the original edge model after the addition of
the non-local coupling~(\ref{eq:coulomb-action}).  To illustrate the
mutual consistency of the two models, we derive the
relationship between the coupling ${\cal K}(\omega)$ in the two
tunneling geometries directly, using only the quadratic action 
${\cal S}_{\rm q}\equiv {\cal S}_0+{\cal S}_1$.

\begin{figure}[htbp]
  \begin{center}
    \leavevmode
    \epsfxsize=0.9\columnwidth
    \epsfbox{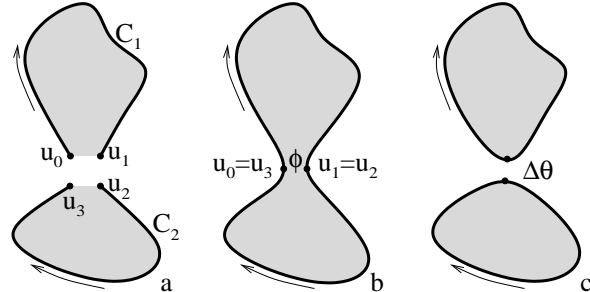}
    \caption{The auxialiary edge configuration (a) is used to calculate
      the quadratic part of the action for the tunneling geometries
      (b), (c).  It is assumed that the short-distance cut-off for the
      long-range potential $V(r)$ is much larger then the scale at
      which the geometries (b) and (c) differ.}
    \label{fig:dual-proof}
  \end{center}
\end{figure}

Consider a field configuration with the boundary conditions fixed as
in Fig.~\ref{fig:dual-proof}a, where $u_i=u_i(\tau)$ are given.
Everywhere on the composite contour $C\equiv C_1+C_2$ the action is
quadratic, and the corresponding Euler-Lagrange equation is linear,
$$
\partial_x\left[i\partial_\tau  u+v(x)\,\partial_x   u+
  {\nu\,e^2\over 2\pi\varepsilon}\int_{C} dy\, V(|{\bf r}_x-{\bf
    r}_y|)\,\partial_y u\right]=0.
$$
The classical solution is uniquely determined by the given values
$u_i(\tau)$ of the fields at the endpoints.  The quadratic
action~(\ref{eq:finite-size-edge-action}), (\ref{eq:coulomb-action}),
evaluated along this classical solution, can be written as
\begin{eqnarray}
  {\cal S}_{\rm q}[u]&=&{\cal G}[u_1\!-\!u_0,\,u_3\!-\!u_2]
  +\int d\tau{(u_1\!-\!u_0)\,i\partial_\tau (u_1\!+\!u_0)\over
  4\pi}\nonumber\\ 
  & & 
  +\int d\tau{(u_3\!-\!u_2)\,i\partial_\tau (u_3\!+\!u_2)\over 4\pi}, 
  \label{eq:formally-evaluated-action}
\end{eqnarray}
where ${\cal G}[a,b]$ is a quadratic, non-local in time, and
generally very complicated functional of its arguments.

The conservation of the total charge 
\begin{equation}
  Q={\sqrt\nu \over 2\pi}(u_3-u_2+u_1-u_0)
  \label{eq:charge-conserv}
\end{equation}
requires that $\varphi\equiv{}u_1-u_0=u_2-u_3$, up to a
time-independent constant. Setting the total charge to zero, we
can write Eq.~(\ref{eq:formally-evaluated-action}) as
\begin{equation}
  \label{eq:formally-evaluated-action-two}
  {\cal S}_{\rm q}[\varphi,\Delta\theta]=
  {\cal G}[\varphi,-\varphi] -{1\over 2\pi}\! \int\!
  d\tau\,{\varphi\,i\partial_\tau \Delta\theta}. 
\end{equation}
where $\Delta\theta\equiv (u_3\!+\!u_2\!-\!u_1\!-\!u_0)/2$.
For the tunneling geometry in Fig.~\ref{fig:dual-proof}b,
Eq.~(\ref{eq:charge-conserv}) implies that the field $u(x,\tau)$ can
be chosen continuous everywhere along the combined edge $C_1+C_2$, 
$\Delta\theta=0$ and hence the effective quadratic action becomes 
$$
S_{\rm q} ={\cal G}[\varphi,\,-\varphi]\equiv {T\over4\pi}\sum_{\omega_n=
  2\pi\,nT} |\omega_n| \,{\cal K}(\omega_n)\,|\varphi_n|^2 ,
$$
where we introduced the coupling ${\cal K}(\omega)$ as in
Eq.~(\ref{eq:effective-model}).

For the tunneling geometry in Fig.~\ref{fig:dual-proof}c, the charges
in upper and lower areas change with time as a result of the
tunneling, and we must keep the field $u(x,\tau)$ discontinuous.  The
corresponding action becomes
$$
\tilde S_{\rm q} = {T\over4\pi}\sum_{n} |\omega_n| \,{\cal
  K}(\omega_n)\,|\varphi_n|^2 +\omega_n
(\bar\varphi_{n}\Delta\theta_n-\Delta\bar\theta_{n}\varphi_n).
$$
(Note that a different choice of $\Delta\theta$, {\em e.g.},
$\Delta\theta=u_3-u_0$ or $\Delta\theta=u_2-u_1$, only changes the
Euclidean Lagrangian by a total time derivative, thus leaving the
action $\tilde{S}_{\rm q}$ invariant.)
The field $\varphi$ can be now trivially integrated out, 
and we arrive at the final form of quadratic action for this geometry, 
\begin{equation}
  \label{eq:dual-effective-action}
  \tilde S_{\rm q} = {T\over4\pi}\sum_{\omega_n}
  |\omega_n| \,\tilde{\cal K}(\omega_n)\,|\Delta\theta_n|^2, 
  \quad \tilde{\cal K}(\omega_n) ={1\over{\cal K}(\omega_n)}.
\end{equation}
This result can be generalized for systems with several junctions,
where the coupling ${\cal K}(\omega)$ is replaced by a matrix, which
is inverted when all junctions are replaced by their
duals\cite{Pryadko-98}.

This simple calculation shows that even in the presence of long-range
interactions the duality between weak tunneling and weak
backscattering for the model described by
Eqns.~(\ref{eq:finite-size-edge-action}), (\ref{eq:coulomb-action}),
(\ref{eq:tunn-action}) coincides with the duality between weak and
strong coupling for the effective tunneling
model~(\ref{eq:effective-model}), independent of the actual geometry
of the system.  The only assumption we made is that the geometries in
Fig.~\ref{fig:dual-proof}b and Fig.~\ref{fig:dual-proof}c should not
differ ``substantially'', that is, the size of a junction near a
saddle point should be sufficiently small ({\em e.g.}, compared with a
short-distance cut-off length, or, at small enough frequencies, with
the wavelength $v/\omega$), so that the Coulomb potential would be the
same in the points $u_0,\ldots,u_3$.

\section{Scale-invariant models}
\label{sec:self-similar}

In the absence of long-range forces ($e^2=0$),
the properties of any system are determined only by the relative
location of the tunneling points along the edges.  If such a system
has only one tunneling point, in the limit where both contours $C_1$
and $C_2$ in Fig.~\ref{fig:dual-proof} become infinite, the system
would not ``know'' the difference between the geometries in
Fig.~\ref{fig:dual-proof}b and Fig.~\ref{fig:dual-proof}c, and the
duality implies that the coupling has a universal self-dual value
${\cal K}(\omega)=1$, independent of the actual geometry of the edges.
Of course, this statement requires that $\omega\,L/v\gg1$, otherwise
one can obtain\cite{Pryadko-98} for Figs.~\ref{fig:dual-proof}b and
\ref{fig:dual-proof}c respectively
\begin{displaymath}
  {\cal K}^{(\ref{fig:dual-proof}b)}=
  \left[{\cal K}^{(\ref{fig:dual-proof}c)}\right]^{-1} 
  =
  {1\over2}\left|\coth\Bigl({\omega L_1\over 2v}\Bigr)
      +\coth\Bigl({\omega L_2\over 2v}\Bigr)\right|,
\end{displaymath}
where $L_i$ is the length of the contour $C_i$, and a uniform edge
velocity $v(x)=v$ is assumed for simplicity.

In the presence of Coulomb interactions, the functional form ${\cal
  K}(\omega)$ has been previously found\cite{Moon-96,Oreg-96,Imura-97}
only for two {\em parallel\/} edges ($\alpha\to 0$ or $\alpha=\pi$ in
Fig.~\ref{fig:angles}), where the translational symmetry of the
quadratic part of the action is restored.
In any other geometry the distance $|x-y|$ measured along the edges,
and the geometrical distance $R_{xy}\equiv |{\bf r}_x-{\bf r}_y|$ in
Eq.~(\ref{eq:coulomb-action}) are no longer equivalent, and an
analytic computation of the average~(\ref{eq:thermal-average}) with
``non-interacting'' quadratic action ${\cal S}_0+{\cal S}_1$ becomes
virtually impossible.

Some simplification can be achieved for an idealized {\sf X}-shaped
geometry (see Fig.~\ref{fig:angles}), which can be also introduced as
the zero-bias limit of the edges in a vicinity of a saddle point with
the opening angle $\alpha$.  For the special case of unscreened
Coulomb potential,
\begin{equation}
  \label{eq:coulomb}
  V(R)=\left({R^2+a^2}\right)^{-1/2},
\end{equation}
the long-range interaction term~(\ref{eq:coulomb-action}) scales the
same way as the local potential (velocity) term in
Eq.~(\ref{eq:edge-action}).  Then, if the edge velocity $v(x)=v$ is
coordinate-independent the action becomes scale invariant for a
sufficiently small short-distance cutoff $a$.  This implies that the
function ${\cal K}_\alpha(\omega)$, for a given opening angle
$\alpha$, can depend on the frequency at most logarithmically.  In
this regime the geometry of the edges and the tunneling properties of
the junction [{\em i.e.}, the function ${\cal K}_\alpha(\omega)$] are
fully determined by the angle $\alpha$ and the dimensionless coupling
constant\cite{Moon-96}
\begin{equation}
  \chi\equiv {\nu\,e^2/(\pi\hbar
    v\varepsilon)}\label{eq:coupling-constant}. 
\end{equation}

The duality discussed in the previous section implies that ${\cal
  K}_{\pi-\alpha}(\omega)={\cal K}^{-1}_{\alpha}(\omega)$, for given
values of the coupling constant $\chi$ and the cut-off scale $a$.
Therefore, in the self-dual geometry at $\alpha=\pi/2$, we expect
$K_{\pi/2}=1$ exactly, independent of the form or the strength of the
interaction potential $V(x)$.

To rewrite more explicitly the general Coulomb
action~(\ref{eq:coulomb-action}) for the infinite geometry in
Fig.~\ref{fig:angles}, let us introduce the coordinate $x$ along each
edge, with the origin at the tunneling point and positive direction to
the right.  Then the charge densities along the top and the bottom
boundaries are respectively
$\rho_1(x)=\sqrt\nu\partial_x\,u_1(x)/(2\pi)$ and
$\rho_2(x)=-\sqrt\nu\partial_x\,u_2(x)/(2\pi)$ (the sign in the second
expression differs because the coordinate is now chosen in the
direction opposite to the edge velocity).  The Coulomb part of the
action becomes
$$
{\cal S}_1= {\chi\over8\pi}\!\int\!
d\tau\!\int_{-\infty}^\infty\!\! dx\,dy \!  \sum_{i,j=1,2}(-1)^{i+j}
\partial_x u_iV_{ij}(x,y)\partial_y u_j,
$$
where the potential $V_{ij}(x,y)\equiv V\biglb(|{\bf r}_{i}(x)-{\bf
  r}_{j}(y)|\bigrb)$ denotes the interaction energy between unit
charges at the points $x$ and $y$ at the edges $i$ and $j$
respectively, and we changed the units of distance: from now on $v=1$.
For symmetric geometries $V_{ij}(x,y)=V_{ij}(y,x)$,
$V_{11}(x,y)=V_{22}(x,y)$, and the obtained expression can be
diagonalized by introducing the symmetric and antisymmetric
combinations $\varphi=u_1-u_2$, $\vartheta=u_1+u_2$.  The quadratic
part (\ref{eq:finite-size-edge-action}), (\ref{eq:coulomb-action}) of the
Euclidean action becomes
\begin{eqnarray}
  \lefteqn{{\cal S}_{\rm q}={T\over8\pi}\sum_n\biggl\{\int dx\left[ 
      2\omega_n\bar\varphi(x)\,\vartheta'_x+|\varphi'_x|^2
      +|\vartheta'_x|^2\right]
    \nonumber} & &  \\
  & &\;\, + {\chi\over2} \int \!dx\!\int
      \!dy\,\left[\bar\varphi'_x\,V_+(x,y)\,\varphi'_y 
    +\bar\vartheta'_x\,V_-(x,y)\,\vartheta'_y\right]\biggr\},
  \label{eq:symmetrized-action}
\end{eqnarray}
where $V_\pm(x,y)\equiv V_{11}(x,y)\pm V_{12}(x,y)$ and the coordinate
integrations are performed along the entire real axis.  Note that the
first term of the integrand is not written as
$\omega_n(\bar\varphi\,\vartheta'_x-\bar\vartheta\,\varphi'_x)$ as
would be expected from the action~(\ref{eq:edge-action}); the
integrand in Eq.~(\ref{eq:symmetrized-action}) differs by a full
spatial derivative, exactly equivalent to the surface term in the
second line of Eq.~(\ref{eq:finite-size-edge-action}).

The interaction potential is always an even function with respect
to simultaneous reflection of both coordinates,
$V_\pm(x,y)=V_\pm(-x,-y)$, and the fields
$\varphi=\varphi_s+\varphi_a$ and $\vartheta=\vartheta_s+\vartheta_a$
can be separated into symmetric ($s$) and antisymmetric ($a$)
components.  The first term of the action~(\ref{eq:symmetrized-action})
couples only the components of two fields with the opposite symmetry:
$\varphi_s$ with $\vartheta_a$, and $\varphi_a$ with $\vartheta_s$.
Since the tunneling term depends on the field
$\varphi(0)=\varphi_s(x=0)$ only, the components $\varphi_a(x)$ and
$\vartheta_s(x)$ decouple and can be integrated out independently of
the value $\varphi(0)$.  In the following, we shall presume that this
symmetrization has been done, and use
\begin{equation}
  \varphi(x)=\varphi(-x),\quad \vartheta(x)=-\vartheta(-x),
  \label{eq:field-symmetry}
\end{equation}
with the indices ``$s$'' and ``$a$'' dropped for convenience.

\subsection{Exactly-solvable example}

To illustrate the properties of the symmetrized
action~(\ref{eq:symmetrized-action}), consider a model problem where
the interaction happens only between the points at equal distance from
the origin,
\begin{eqnarray*}
  (\chi/2)\, V_{11}(x,y)&=&v_0\delta(x-y)+v_1\delta(x+y),\\ 
  (\chi/2)\, V_{12}(x,y)&=&v_2\delta(x-y)+v_3\delta(x+y), 
\end{eqnarray*}
where the velocity $v_0$ (measured in units of bare velocity $v$)
denotes the strength of additional interaction at the same edge, $v_1$
and $v_2$ denote the interaction between the neighboring edges
(left--right and top--bottom), while $v_3$ denotes the interaction
between the points at the opposing edges.  (Physically, this set of
interactions corresponds to four locally-interacting chiral edges,
running along the surface of a semi-infinite cylinder and meeting in
the tunneling point at its near end).

With interaction of this simple form we can use the symmetry
properties~(\ref{eq:field-symmetry}), and the quadratic
action~(\ref{eq:symmetrized-action}) becomes entirely {\em local},
$$
{\cal S}_{\rm q}={T\over8\pi}\sum_n\int dx\left[
  2\omega_n\bar\varphi(x)\,\partial_x\vartheta+ v_\varphi
  |\partial_x\varphi|^2+ v_\vartheta |\partial_x\vartheta|^2 \right],
$$
with $v_{\varphi,\vartheta}=1+v_0-v_3\pm (v_1-v_2)$.  Now the field
$\vartheta(x)$ can be trivially integrated out, and
Eq.~(\ref{eq:thermal-average}) gives
$$
{\cal K}^{-1}(\omega_n)={1+v_0-v_3+ (v_1-v_2)\over 1+v_0-v_3-
  (v_1-v_2)}. 
$$
Clearly, under interchange $v_1\leftrightarrow v_2$ this expression
goes to its inverse according to the duality relation derived in
Sec.~\ref{sec:duality}, and ${\cal K}(\omega_n)=1$ for the self-dual
case $v_1=v_2$ where all edges are equivalent.

\subsection{Coulomb interactions near a saddle point}
\label{sec:coulomb-wedge}

Now let us consider more realistic long-distance interactions in the
edge geometry shown in Fig.~\ref{fig:angles}.  We write the intra- and
inter-edge interaction potentials
\begin{eqnarray*}
  V_{11}(x,y)&=&\tf(xy)V(x-y)  +\tf(-xy)V(R_\alpha),\\
  V_{12}(x,y)&=&\tf(xy)V(R_\alpha)  +\tf(-xy)V(x-y),
\end{eqnarray*}
where the bulk distance $R_\alpha\equiv
({x^2+y^2-2xy\cos\alpha})^{1/2}$, $\tf(x)$ is the usual step function,
$\tf(x)=1$ for $x>0$ and $\tf(x)=0$ otherwise, and $V(x)$ is, {\em
  e.g.}, the Coulomb potential~(\ref{eq:coulomb}).  The resulting
effective action has the form~(\ref{eq:symmetrized-action}), with
\begin{eqnarray}
  \label{eq:v-plus}
  V_+&=&V(x-y) +V(R_\alpha),\\
  V_-&=&\left[V(x-y) -V(R_\alpha)\right]\sgn (xy). 
  \label{eq:v-minus}
\end{eqnarray}

In the limit $\alpha=0$, $R_\alpha=|x-y|$, the antisym\-met\-ric part
of the potential vanishes, $V_-(x,y)=0$, while $V_+(x,y)=2V(x-y)$, and
we obtain the usual translationally-invariant action for two parallel
edges.  Integrating out the field $\vartheta$ and diagonalizing the
remaining part of the action with the help of Fourrier
transformation, we use Eq.~(\ref{eq:thermal-average}) to calculate the
coupling,
\begin{equation}
  \label{eq:coupling-zero-angle}
  {\cal K}_{\alpha=0}^{-1}(\bar\omega)
  ={2\bar\omega\over\pi}\int_0^\infty {d\zeta\over
    \bar\omega^2+\zeta^2\,[1+2\chi\,K_0(\zeta)] }, 
\end{equation}
where the Fourrier-transformed Coulomb potential~(\ref{eq:coulomb}),
$V(\zeta)=2K_0(\zeta)$, is expressed in terms of the modified Bessel
function $K_0$, and the reduced frequency $\bar\omega=a\omega/v$.
Performing the integration with logarithmic accuracy, we obtain, in
agreement with Refs.~\CITE{Moon-96,Oreg-96,Imura-97}
\begin{equation}
  {\cal K}_{\alpha=0}= \left[1+2\chi\ln\left({
  2\sqrt{2\chi}\,e^{-\gamma}\over\bar\omega }\right)\right]^{1/2}
      +{\cal O}\bigl(|\ln \bar\omega|^{-1/2}\bigr),  
  \label{eq:alpha0}
\end{equation}
where $\gamma\approx 0.577$ is the Euler constant.

The case $\alpha=\pi$ corresponds to two semi-infinite non-chiral
Luttinger liquids connected by a tunneling point ($\alpha\to\pi$ in
Fig.~\ref{fig:angles}); by duality we
expect\cite{Imura-97} ${\cal K}_{\alpha=\pi} =1/{\cal K}_{\alpha=0}$.
This expression is proved again, specifically for this geometry, in
Appendix~\ref{sec:appendix-pi}.

We argued that in the self-dual case $\alpha=\pi/2$, ${\cal
  K}(\omega)=1$ identically, independently of the properties of the
potential $V(x)$, as long as it is appropriately regularized at short
distances.  We have also constructed a direct analytical solution for
this case.  The major simplification comes from an observation that
the potential $V(R_{\pi/2})= V(\sqrt{x^2+y^2})$ is a symmetric
function of $x$ and $y$ independently; the corresponding contribution
vanishes from the action~(\ref{eq:symmetrized-action}) by the
symmetry~(\ref{eq:field-symmetry}).  As a result, only the potentials
$V(x\pm y)$ with the distance measured along the edge enter the
extremum equations, and these equations can be solved exactly using
the Wiener-Hopf method, as detailed in Appendix~\ref{sec:wiener-hopf}.
This direct solution confirms the universal result ${\cal
  K}_{\alpha=\pi/2}=1$.  In addition, the explicitly found extremum
configuration of the fields $\varphi(x)$, $\vartheta(x)$ is used to
get a perturbative expression for ${\cal K}_\alpha(\omega)$ near the
self-dual point $\alpha_0=\pi/2$.  This yields (see Appendix B)
\begin{equation}
  \label{eq:perturbation}
  K_\alpha(\omega\to0)\approx 1+{\cal N}(\chi)\,\chi\cos\alpha, 
\quad |\cos\alpha|\ll 1,
\end{equation}
where ${\cal N}(\chi)$ is {\em independent of} $\omega$.  In the limit
of weak Coulomb interactions, ${\cal N}(\chi\to0)\approx1.51$, while
${\cal N}(\chi=1.0)\approx0.21$.

To get a handle on the dependence of the coupling ${\cal
  K}_\alpha(\omega)$ on the parameters and the cut-off scales, we have
also evaluated the average~(\ref{eq:thermal-average}) numerically for
the quadratic action~(\ref{eq:symmetrized-action}) with the Coulomb
potential~(\ref{eq:coulomb}) at different frequencies $\omega$, and
for different values of the angle $\alpha$ and the dimensionless
coupling constant $\chi$.

To perform this calculation we wrote a discretized version of the
quadratic action~(\ref{eq:symmetrized-action}) in terms of lattice
values $\varphi(x_n)$ and $\vartheta'(x_n)$, $0<n<N-1$, and then
integrated out the values of the fields away from the origin, which
only required inverting two $N\times N$ matrices.  In addition to the
cut-off distance $a$ in Eq.~(\ref{eq:coulomb}), the discretization
involved two explicit cut-off scales: the total system size $L$ and
the lattice grid size $h=L/N$.  The calculations were performed in the
regime $h\ll a\ll L$; the results are independent of these cut-off
scales in the frequency range $h\ll v/\omega\ll L$.  These
inequalities substantially limited the dynamical range where the
results are accurate.

\begin{figure}[htbp]
  \begin{center}
    \leavevmode
    \epsfxsize=\columnwidth
    \epsfbox{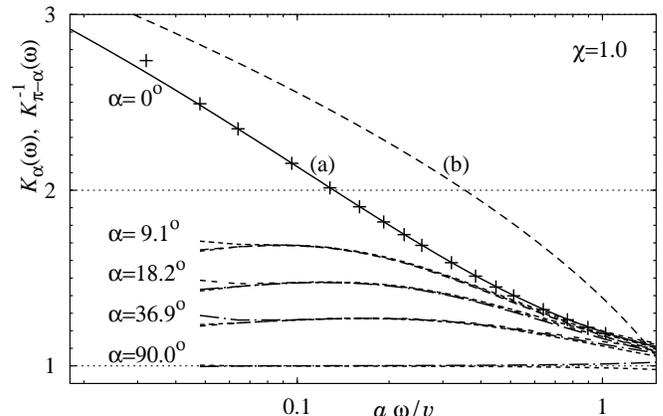}
    \caption{Superimposed values of ${\cal K}_\alpha(\bar\omega)$, ${\cal
        K}_{\pi-\alpha}^{-1}(\bar\omega)$ calculated numerically for
      the Coulomb potential~(\protect\ref{eq:coulomb}) at different
      values of the opening angle $\alpha$, with $L=20$ and the
      lattice size $N=1600$.  The self-duality of the original
      action~(\protect\ref{eq:symmetrized-action}) is violated by the
      finite system size $L$ at small $a\omega/v$, and by the
      discreteness of the lattice spacing at large $a\omega/v$.
      Pluses show the numerical data at $\alpha=0$, while the lines
      (a) and (b) respectively correspond to
      Eqns.~(\protect\ref{eq:coupling-zero-angle}) and
      (\protect\ref{eq:alpha0}) in the text.}
    \label{fig:K-frequency}
  \end{center}
\end{figure}

\begin{figure}[htbp]
  \begin{center}
    \leavevmode
    \epsfxsize=\columnwidth
    \epsfbox{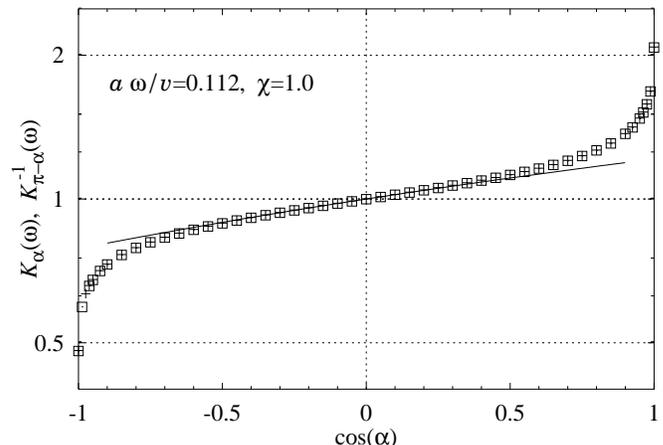}
    \caption{Duality for junctions in Fig.~\protect\ref{fig:angles}
      illustrated nu\-mer\-i\-cal\-ly.  Boxes and Pluses show the
      superimposed values ${\cal K}_\alpha(\omega)$ and ${\cal
        K}_{\pi-\alpha}(\omega)$, calculated with the Coulomb
      potential~(\protect\ref{eq:coulomb}) at the lattice size
      $N=1600$, $L=20$.  Solid line with the slope ${\cal
      N}(\chi=1)\approx 0.212$ is a fit to the data.}
    \label{fig:K-angle}
  \end{center}
\end{figure}

Typical results of the calculations are illustrated in
Fig.~\ref{fig:K-frequency} and Fig.~\ref{fig:K-angle}.  The curves in
Fig.~\ref{fig:K-frequency} with marked values of $\cos(\alpha)$ show
superimposed values ${\cal K}_\alpha(\bar\omega)$, ${\cal
  K}^{-1}_{\pi-\alpha}(\bar\omega)$ calculated with the lattice size
$N=1600$, for cut-off parameters $a=0.05$, $0.1$.  The deviation
betwen the curves shows that our discretization violated the
self-duality of the problem at both large and small cutoff scales.
Nevertheless, as illustrated in Fig.~\ref{fig:K-angle}, the
self-duality holds with a very good numerical accuracy near the middle
of the dynamical range, $a\,\omega/v\sim 0.1$.

As indicated by finite-size scaling analysis of our data (not shown),
at small enough $\omega$, ${\cal K}_\alpha(\omega)$ saturates to a
frequency-independent value in the range $0<\alpha<\pi$.  This
behavior is consistent with the small-angle
expansion~(\ref{eq:perturbation}).  In addition,
Fig.~\ref{fig:K-angle} indicates that Eq.~(\ref{eq:perturbation})
provides a good approximation to ${\cal K}_\alpha(\omega)$ in a rather
wide range of $\alpha$.  For small $\alpha\ll1$, as the frequency is
reduced, the numerical values ${\cal K}_\alpha(\omega)$ seem to
closely follow the logarithmically divergent line (\ref{eq:alpha0}),
but eventually cross over to a constant value ${\cal
  K}_\alpha(\omega=0,\chi)$, which (logarithmically) depends on the
angle and the cut-off scale $a$.

\subsection{Coulomb interactions in the cleaved edge geometry}
\label{sec:Grayson}

Here we consider the effect of long-range interactions in the
cleaved-edge geometry\cite{Chang-96,Grayson-98}, where the tunneling
happens between a three-dimensional metal and the edge of a 2DEG,
located in the plane perpendicular to the surface of the metal.  It is
believed that the tunneling in these experiments is dominated by
localized ``hot'' spots or impurities.  Chamon and
Fradkin\cite{Chamon-97} demontstated that in the absence of
interactions, a point contact between a 3D metal and a QH edge with
the filling fraction $\nu$ is equivalent to a point tunneling junction
between such an edge and an ideal non-interacting $\nu=1$ edge;
furthermore, they mapped this latter problem to that of tunneling
between two identical edges with filling fractions
$\nu_*=2\nu/(1+\nu)$.

\begin{figure}[htbp]
  \begin{center}
    \leavevmode
    \epsfxsize=\columnwidth
    \epsfbox{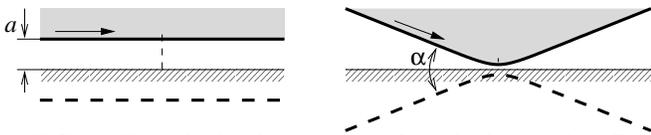}
    \caption{Two idealized geometries for calculating the effect of
      Coulomb interactions in cleaved-edge experiments.  Edge
      quasiparticles interact with image charges induced on the
      metallic surface.}
    \label{fig:cleaved}
  \end{center}
\end{figure}

The effect of the Coulomb interaction in this setup is limited to the
chiral Luttinger liquid, the ``real'' quantum Hall edge, while the
Fermi-liquid nature of quasiparticles in the metal imply that they
remain non-interacting for the purposes of tunneling measurements.
The metallic surface only provides additional screening charges, which
modify the form of the interaction potential $V\left(|{\bf r}_x-{\bf
    r}_y|\right)$.  Assuming characteristic frequencies at the edge
are small compared with the plasma frequency of electrons in metal
(which is always true for a good metal), the retardation can be
neglected, and the modified interaction potential is obtained simply
by adding the appropriate image charges.

The quadratic part of the action for the translationally-invariant
geometry shown in the left part of Fig.~\ref{fig:cleaved} 
({\em i.e}., the case $\alpha=0$) is obtained
by combining Eq.~(\ref{eq:finite-size-edge-action}) with the Coulomb
energy
\begin{equation}
  \label{eq:cleaved-action-coulomb}
  {\cal S}_1= 
  {\chi\over 8\pi}\int_0^\beta d\tau\int_{-\infty}^\infty \!\!dx\,dy
  \,\partial_x u\,\hat V(x-y)\, \partial_y u ,
\end{equation}
where $\hat V(x)\equiv V(x)-V(\sqrt{x^2+4a^2})$ is corrected for the
image potential, and the units of length are again chosen so that the
edge velocity $v=1$.  Because we work with the chiral field now, the
surface term in the second line of
Eq.~(\ref{eq:finite-size-edge-action}) is absolutely essential even in
an infinite geometry.  To properly account for this term, we formally
separate the field $u=\phi+\theta$ into its symmetric
$\phi(x,\tau)=\phi(-x,\tau)$ and antisymmetric
$\theta(x,\tau)=-\theta(-x,\tau)$ compontents; then the surface term
can be absorbed after an integration by parts, and the
action~(\ref{eq:finite-size-edge-action}) becomes
\begin{equation}
  \label{eq:cleaved-symmetrized-quadratic}
  {\cal S}_0\!=\!{1\over 4\pi}\! \int
  \! d\tau\!\int_{-\infty}^\infty\!\! dx \left[
    2i\partial_\tau\phi\,\partial_x\theta+ (\partial_x\phi)^2+
    (\partial_x\theta)^2\right].
\end{equation}
This transformation is equivalent to ``folding'' the chiral edge in
half, which produces two non-chiral fields defined on a semiaxis, and
simultaneously eliminates the zero mode and associated subtleties.
The translationally invariant action can be now diagonalized by a
Fourrier transformation, and, after integrating out the fluctuations
away from the origin, we obtain the single-edge contribution to the
quadratic part of the effective action,
\begin{equation}
  \label{eq:half-action}
  {\cal S}^{(1)}_{\rm q}={T\over 2\pi}\sum_n |\omega_n| \,\hat{\cal 
    K}(\omega_n)\,|\phi_1|^2,
\end{equation}
where $\phi_1\equiv u(0)=\phi(0)$ by definition, and
\begin{eqnarray}
  \label{eq:cleaved-parallel-K}
  \hat{\cal K}^{-1}(\omega)&=&{2|\omega|\over\pi}\!\int_0^\infty\!\!
  {dk\,Z(k)\over  \omega^2+k^2\,Z^2(k)}, \\  \nonumber 
  Z(k)&=&1+{\chi\over2}\,\hat V(k).
\end{eqnarray}
The argument\cite{Chamon-97} that a point contact with a metal is
equivalent to that with a non-interacting $\nu=1$ edge holds
independently of the interactions affecting the ``real'' edge. 
Therefore, the full effective action can be written as
\begin{equation}
  \label{eq:full-asymmetric-action}
  {\cal S}={T\over2\pi}\sum_n |\omega_n|
  \left({ \hat{\cal K}\,|\phi_1|^2
      \!  +|\phi_2|^2}\right) 
  \!+\!\int\! d\tau\,\re\,\lambda\,e^{i(g \phi_1-\phi_2)},
\end{equation}
where we used $\hat{\cal K}=1$ for the auxiliary $\nu=1$ edge.  The
canonical form~(\ref{eq:effective-model}) of the tunneling action can
be obtained by introducing the tunneling degree of freedom, $\varphi=g
\phi_1-\phi_2$, with the corresponding effective coupling ${\cal
  K}_{\rm eff}$ calculated, {\em e.g.}, using the average as in
Eq.~(\ref{eq:thermal-average}).  As before, the resulting model
describes an overdamped particle in a washboard potential; the
corresponding non-Ohmic ``friction'' coefficient
\begin{equation}
  \kappa_{\rm eff}(\omega)\equiv{{\cal K}_{\rm eff}\over g_{\rm
      eff}^2}={2\over g^2/\hat{\cal K}(\omega)+1}. 
  \label{eq:friction-cleaved}
\end{equation}
In the non-interacting limit $\hat{\cal K}(\omega)=1$ this
expression safely goes into the result\cite{Chamon-97} obtained by a
different method.  

Notice that the long-distance part of the Coulomb potential
$\hat V$ in Eq.~(\ref{eq:cleaved-action-coulomb}) is screened by the
metallic surface.  Then, at sufficiently small frequencies,
$a\omega\ll v_{\rm r}\equiv Z(0)\,v$, the momentum dependence of the
coefficient $Z(k)$ can be ignored, and the
integral~(\ref{eq:cleaved-parallel-K}) gives precisely the
non-interacting coupling, $\hat{\cal K}=1$.  This is not at all
surprising, since the interaction happens within a single chiral edge,
and its long-range part (most dangerous at small frequencies) is
screened.  As usual\cite{Volkov-88}, the only effect of the additional
interaction in this chiral system is the velocity renormalization,
$v\to v_{\rm r}$.

The translational symmetry is lost for the ``wedge'' geometry shown in
the right part of Fig.~\ref{fig:cleaved}.  The Coulomb part of the
corresponding action can be written in the
form~(\ref{eq:cleaved-action-coulomb}) with the potential $\hat
V(x-y)\to
V_-(x,y)$ given by Eq.~(\ref{eq:v-minus}).  In the limit $\alpha\to0$,
the potential $V_-(x,y)$ vanishes identically, and hence 
$\hat{\cal K}(\omega)=1$ in this case as well.

At general values of $\alpha$ we again use the ``folding'' trick by
introducing symmetric and antisymmetric variables $\phi$, $\theta$.
Up to an overall coefficient, the resulting action looks like
Eq.~(\ref{eq:symmetrized-action}), with the exception that both
components $\phi$ and $\theta$ couple with the {\em same\/} potential
$V_-(x,y)$.  The most prominent difference is that at $\alpha=\pi/2$
the symmetry no longer leads to a cancellation of the part
$V(\sqrt{x^2+y^2})$ of the total potential, and the effect of the
long-distance interactions is no longer trivial, $\hat{\cal
  K}_{\pi/2}(\omega)\neq1$.  Again, this comes as no surprise, since
there is no self-duality in this geometry.

Finally, in the limiting case $\alpha=\pi$, the potential $V_-(x,y)$
becomes an even function of each argument; as a result, the coupling
with the symmetric field $\phi$ (antisymmetric derivative
$\partial_x\phi$) vanishes by symmetry.  Up to an overall coefficient,
the resulting action is identical to that considered in
Appendix~\ref{sec:appendix-pi}, and we obtain [note that the extra
  coefficient was already accounted for in the corresponding effective
  action, {\em cf}.\ Eqns.~(\ref{eq:half-action})
  and~(\ref{eq:effective-model})],
$$
\hat{\cal K}_{\alpha=\pi}(\omega)={\cal K}_{\alpha=\pi}(\omega)
={2|\omega|\over \pi}\!\int_{0}^\infty\!\! {dk\over \omega^2+k^2
  \biglb(1+\chi\,V(k)\bigrb)}.
$$
This result is quite intuitive: metallic screening becomes non-effective
in the case where a wire is perpendicular to the conducting surface.

Our calculations imply that the tunneling exponent is modified by
the Coulomb interaction 
only if the edge is bent near the tunneling point.  In an ideal
sample, the edge runs along a straight line parallel to the surface of
the metal, and long-range interactions do not modify the tunneling
exponents.  In any real sample, however, imperfections near the
tunneling point always reduce the effective coupling ${\cal
  K}(\omega)$, or, equivalently, systematically {\em increase\/} the
tunneling exponent in Eq.~(\ref{eq:lameff}).  Nevertheless, we do not
believe this effect would be sufficient to explain a $10\%$ increase
of the tunneling exponent observed\cite{Grayson-98} by Grayson {\em et
  al}.\ near $\nu=1$:  
cleaved-edge samples are characterized by sharp confinement and large
drift velocities, meaning that the corresponding dimensionless
coupling constant $\chi$ [see Eq.~(\ref{eq:coupling-constant})] is
small.

\section{Discussion}
\label{sec:discussion}

We have shown that the effect of long-range interactions on transport
through a QH tunneling junction depends crucially on its geometry.  In
particular, in a self-similar {\sf X}-shaped junction (see
Fig.~\ref{fig:angles}) characterized by an opening angle $\alpha$,
unscreened Coulomb interactions only renormalize the effective
Luttinger-liquid exponent,
$$
g_\star^2=g^2/{\cal K}_\alpha(\omega=0,\,\chi),
$$
where $g^2=1/\nu$ for electron tunneling between the edges of 2DEGs with
Laughlin fractions $\nu$.  Therefore, the renormalized exponent
depends non-universally on the angle $\alpha$ and the dimensionless
Coulomb interaction strength $\chi$.

This implies that the system should exhibit a zero-temperature
delocalization transition at a critical angle characterized by
$g_\star^2=1$.  This is in contrast with the transport properties
expected in the absence of long range interactions, which are
exclusively determined by the filling fraction $\nu$.  For integer QH
systems with $\nu=1$, the transition always corresponds to a self-dual
geometry, {\em i.e.}, $\alpha_c=\pi/2$, independently of the details
of the interaction.  In fractional QH constrictions, however, the
transition (if any) occurs at a non-universal critical angle
$\alpha_c<\pi/2$, such that ${\cal K}_{\alpha_c}(0,\chi)=\nu^{-1}$.

Properties of all charge transfer processes through the junction are
defined by the parameter $g_\star$ in the effective
action~(\ref{eq:formally-evaluated-action-two}), which determines the
tunneling exponents\cite{Wen-91B,Kane-Fisher-Tunnel} [see
  Eq.~(\ref{eq:lameff})], the form of the non-linear $I$--$V$
curve\cite{Weiss-Exact,Fendley-95B}, as well as the tunneling
noise~\cite{Kane-94A,Fendley-95C,QHpers-book}.  In the limit of weak
tunneling, the quantization of transferred charge is ultimately
determined by gauge invariance, and a shot noise measurement would
show current transferred by unit charges.  However, the shot noise
measured in the opposite, strongly coupled limit (reached, {\em e.g.},
by driving a large tunneling current through the junction), is
set\cite{Sandler-noise} by the instanton charge for the effective
tunneling action~(\ref{eq:formally-evaluated-action-two}).  The value
of this charge is determined solely by the value of $g_\star$.  Hence,
in this regime a noise measurement would show a non-universal charge
$$
e_\star/e=1/g^2_\star=\nu\,{\cal K}_\alpha(0,\,\chi),
$$ 
clearly an interaction effect.  

The described situation applies to ideal systems without screening.
More realistically, Coulomb interactions are screened at some finite
length $\xi$.  Then, for a junction with finite opening angle,
$|\cos\alpha|<1$, the correction to tunneling exponents always
vanishes in the static limit, ${\cal K}^{\rm (scr)}_\alpha(0)=1$, even
though it may be significant at larger frequencies, $\omega\gtrsim
v/\xi$ (this corresponds to a temperature $0.1$~K for $\xi=1\;\,\mu$m
and $v=10^{-7}$~cm/s).  Consequently, a system at a fractional $\nu$
with originally metallic behavior would eventually localize at small
enough temperatures.\cite{Moon-96,Imura-97} Contrarily, the
interaction-induced flow in an integer junction would gradually stop
without changing its direction.

For an {\sf X}-shaped junction with a given opening angle $\alpha$,
the magnitude of the renormalization parameter ${\cal
  K}_\alpha(0,\chi)$ is determined by the value of the dimensionless
Coulomb interaction constant~(\ref{eq:coupling-constant}), which, in
turn, is defined by the edge wave (drift) velocity.  For cleaved-edge
samples, edge magnetoplasmon velocities have been
measured\cite{velocity-cleaved} by Ashoori {\em et al.}, yielding
$v\sim 10^8$~cm/s which corresponds to $\chi\sim 0.05$.  On the other
hand, edge electric fields equivalent to drift velocities as small as
$v\sim 10^6$~cm/s have been measured by Maasilta and
Goldman\cite{Maasilta-97}, who analyzed discrete energy levels of a
quantum antidot.  This value of velocity results in a relatively large
coupling constant value $\chi\sim5$.

We must point out, however, that our discussion of Coulomb interaction
effects was based on a single-mode sharp edge, which implies large
confining electric fields of order ${\cal E}\sim E_g/(e\ell)$, where
$E_g$ is the energy gap associated with the incompressible QH state,
and $\ell$ is the magnetic length.  Using the drift velocity $v=c
{\cal E}/B$, we obtain
\begin{equation}
  \chi={\nu \,e^2\over \pi\varepsilon\hbar v} 
  \sim\left({\nu\,e^2\over \pi\varepsilon\ell}\right)\,E_g^{-1}, 
\end{equation}
which, for a typical QH sample, leads to $\chi\lesssim 1$.  Samples
with much larger values of the Coulomb coupling are likely to have a
tendency to edge reconstruction.  This would lead to additional
polarization at the edge due to neutral modes, and, consequently, a
partial screening of Coulomb interaction.

Therefore, to observe the predicted effects, samples with
well-defined, but not too sharp edges are necessary.  This excludes
the cleaved-edge samples (where the drift velocity $v$ is large), as
well as the samples with electrostatically-defined geometry (where
confinement tends to be soft).  The best choice would therefore be a
Hall bar with lithographically defined {\sf X}-shaped constriction and
a narrow local gate to fine-tune the tunneling.  For a given base
temperature $T$, the linear size of the constriction should be at
least of order $\xi\sim\hbar v/T$, {\em i.e.}, approaching a
millimeter scale for millikelvin temperature range.  Tunneling
junctions with small opening angles will give larger values of ${\cal
  K}_\alpha$ [in principle, limited only by the
logarithm~(\ref{eq:alpha0}), divergent at small frequencies].
However, as illustrated in Fig.~\ref{fig:K-frequency}, for such
junctions the renormalized Luttinger parameter $g_\star^2$ is more
likely to retain some frequency (temperature) dependence, which would
modify the measured exponents.

\acknowledgements

We gratefully acknowledge useful discussions with C.~de~C.~Chamon,
M.~Fogler, C.~Glattli, E.~Gwinn, S.~A.~Kivelson, D.-H.~Lee,
Z.~Nussinov, S.~L.~Sondhi, and X.-G.~Wen.

A.A.\ acknowledges support of the Israeli Science Foundation and the
fund for Promotion of Research in the Technion.  E.S.\ acknowledges
support by grant no. 96--00294 from the United States--Israel
Binational Science Foundation (BSF), Jerusalem, Israel.  L.P.P.\ was
supported in part by DOE Grant No.\ DE-FG02-90ER40542.
 
\appendix

\section{Coupling at $\alpha=\pi$.}
\label{sec:appendix-pi}

Here we derive the form of the coupling ${\cal K}(\omega)$ for the
saddle-point geometry shown in Fig.~\ref{fig:angles} in the special
limit $\alpha=\pi$, which corresponds to two vertical semi-infinite
wires connected by a single tunneling point.  In this case the
distance $R_\alpha=|x+y|$, and the contribution of the symmetric
potential $V_+(x,y)=V(x-y)+V(x+y)$ to
Eq.~(\ref{eq:symmetrized-action}) vanishes by
symmetry~(\ref{eq:field-symmetry}), so that only the part
$V_-(x,y)=[V(x-y)-V(x+y)]\sgn (xy)$ remains.  The symmetry of the
derivative $\partial_x\vartheta$ implies that both parts of the
potential $V_-$ give identical contribution, and the quadratic part of
the action~(\ref{eq:symmetrized-action}) can be written as
\begin{eqnarray}
  \lefteqn{{\cal S}_{\rm q}
      ={T\over8\pi}\sum_n\biggl\{\int_{-\infty}^\infty dx\left[  
      2\omega_n\bar\varphi(x)\,\partial_x\vartheta
      +|\partial_x\varphi|^2 
      +|\partial_x\vartheta|^2\right]
  \nonumber} & &  \\
  & &\quad +  {\chi} \int_{-\infty}^\infty dx
  \,dy\,\left[\partial_x\bar\vartheta\,V(x-y)\,\partial_y\vartheta\,{\rm
      sgn}(xy) 
      \right]\biggr\}. 
  \label{eq:wedge-action-pi}
\end{eqnarray}
Unlike the case $\alpha=0$, the non-local interaction in the second
line cannot be diagonalized by a simple Fourrier transformation; we
need to get rid of the sign function first.  Naively, this could be
done by multiplying both $\varphi(x)$ and $\vartheta(x)$ by ${\rm
  sgn}(x)$.  However, since $\varphi(0)\neq0$, the function
$\varphi(x)\sgn (x)$ would not be continuous at the origin, so
that spurious $\delta$-functions may be generated.  Instead, we define
auxiliary continuous functions $u(x)$, $g(x)$, so that
$$
\varphi(x)=\varphi(0)+\sgn (x)\,u(x),\quad
g(x)=\vartheta(\infty)-\sgn (x)\,\vartheta(x),
$$
and $u(0)=g(\infty)=0$.  After integrating out the field $u(x)$,
the effective action becomes
\begin{eqnarray*}
  {\cal S}_{\rm q}
  &=&{T\over8\pi}\sum_n\biggl\{
  -4\omega_n\bar\varphi(0)\,g(0)\\  & & 
  \quad+\int {dk\over2\pi} |g_k|^2
  \left[    \omega_n^2+k^2\biglb(1+\chi
      \,V(k)  \bigrb)
    \right]\biggr\}. 
\end{eqnarray*}
In the first term here we substitute $g(0)=\int {dk\,g_k/(2\pi)}$
in terms of the Fourrier-transformed field $g_k$, integrate this field
out, and obtain the effective action for the the field $\varphi(0)$
alone,
$$
{\cal S}_{\rm q} ={T\over4\pi}\sum_n \omega_n \,|\varphi(0)|^2
\left[ {2\omega_n\over \pi}\!\int_{0}^\infty\!\! {dk\over
    \omega_n^2+k^2 \biglb(1+\chi\,V(k)\bigrb)}\right];
$$
comparing the result with the general form of the effective
action~(\ref{eq:effective-model}), and the
result~(\ref{eq:coupling-zero-angle}) for $\alpha=0$, we conclude that
$$
{\cal K}_{\alpha=0}(\omega_n)\, {\cal K}_{\alpha=\pi}(\omega_n)=1
$$
exactly, independent of the form of the potential $V(x)$. 

\section{Self-dual tunneling junction, $\alpha=\pi/2$}
\label{sec:wiener-hopf}
\subsection{General Wiener-Hopf solution.} 

Here we give a direct solution of the extremum equations for the
self-dual case $\alpha=\pi/2$.  This solution gives the coupling
${\cal K}_{\pi/2}=1$ directly, without utilizing the self-duality of
the problem. In addition, it allows us to calculate $K_\alpha$
  perturbatively for small values of $|\cos(\alpha)|\ll 1$.

Begin with the Euclidean action~(\ref{eq:symmetrized-action}) at
$\alpha=\pi/2$, 
\begin{eqnarray}
  \label{eq:wedge-action-half-pi}
  {\cal S}_{\rm q}
  & =& {T\over8\pi}\sum_n\int dx\biggl\{
  2\omega_n\bar\varphi(x)\,\vartheta'_x
  +\int dy\,\bar\varphi'_x\,Z(x-y)\,\varphi'_y
  \nonumber \\
  & & \qquad+\int dy\, \bar\vartheta'_x\,Z(x-y)\sgn (xy)\,\vartheta'_y
  \biggr\},
\end{eqnarray}
where the total potential
\begin{equation}
  Z(x-y)=\delta(x-y)+{\chi\over2}\,V(x-y); \label{eq:total-potential}
\end{equation}
note that due to the symmetry~(\ref{eq:field-symmetry}), the
contribution from the part of the potential with geometrical distance,
$V(R_{\pi/2})=V(\sqrt{x^2+y^2})$, was cancelled.  The Euler-Lagrange
equations (valid at $x\neq0$, where the non-linear tunneling term
gives no contribution) are
\begin{eqnarray}
  \label{eq:EL-phi}
  \omega\partial_x\vartheta&-&\partial_x\int_{-\infty}^\infty \!dy \,
  Z(x-y)\,\partial_y\varphi =0,\\
  \label{eq:EL-theta}
  \omega\partial_x\varphi&-&\partial_x \int_{-\infty}^\infty\! dy \,
  Z(x-y)\sgn (xy)\,\partial_y\vartheta=0.
\end{eqnarray}
We assume that both fields are continuous everywhere, and that
$\varphi(x)$ and $\partial_x\vartheta(x)$ vanish at infinity.
Multiplying the first of the obtained equations by $\bar\varphi(x)$,
the second by $\bar\vartheta(x)$, and subtracting the results from the
integrand in the action~(\ref{eq:wedge-action-half-pi}), with the help
of the definition~(\ref{eq:total-potential}) we obtain
\begin{eqnarray}
  \lefteqn{{\cal S}_{\rm q}= {T\over8\pi}\sum_n\!\int\!\!
    dx\,\partial_x\biggl[
    \omega_n\bar\varphi\,\vartheta\!+\!\bar\varphi(x)\!\int\!\! dy
    \,Z(x-y)\,\partial_y\varphi}\nonumber\\
  & & \qquad+ \bar\vartheta(x) \int\! \!dy \,Z(x-y)
  \sgn(xy)\,\partial_y\vartheta\biggr]
  \nonumber\\
  & =& - {T\over8\pi}\sum_n\bar\varphi(0)\,\Delta\varphi'_0, \quad
  \Delta\varphi'_0\equiv \varphi'(0_+)-\varphi'(0_-),
  \label{eq:effective-action-evaluated}
\end{eqnarray}
where the integration was performed over the entire axis excluding the
point $x=0$.  The Euler-Lagrange equations~(\ref{eq:EL-phi}),
(\ref{eq:EL-theta}) can be simplified by defining linear combinations
(symmetric with respect to $x$) 
\begin{equation}
  A,B(x)=[\varphi(x)\pm\vartheta(x)\sgn(x)]/2,
  \label{eq:a-b-defined}
\end{equation}
then, multiplying Eq.~(\ref{eq:EL-theta}) by $\sgn(x)$ and taking
symmetric and antisymmetric combinations of the result with
Eq.~(\ref{eq:EL-phi}), we obtain at $x\neq0$
\begin{equation}
  \label{eq:EL-A}
  \omega\sgn(x)\partial_x A-\partial_x\int_{-\infty}^\infty \!dy \,
  Z(x-y)\,\partial_y A =0,
\end{equation}
and an identical equation (up to the substitution $\omega\to-\omega$) 
for $B(x)$.  We integrate, keeping in mind that Eq.~(\ref{eq:EL-A}) is
valid for $x\neq0$,
\begin{equation}
  \label{eq:integrated-EL-A}
  \omega A\sgn(x)-\int_{-\infty}^\infty \!dy \, Z(x-y)\,\partial_y A
  =C_a\sgn(x),
\end{equation}
where the integration constants in the intervals $x<0$ and $x>0$ were
related using the symmetry $A(x)=A(-x)$.  The value of the constant
$C_a$ is determined by the boundary conditions; using the
definition~(\ref{eq:a-b-defined}) we obtain
\begin{equation}
  \label{eq:discontinuity-a}
  2C_a=  \omega\,\varphi(0)-\varphi'(0_+)-\vartheta'(0)
  =\omega\,\vartheta(\infty). 
\end{equation}
Similarly, the integration of the corresponding equation for the
function $B(x)$ yields
\begin{equation}
  \label{eq:discontinuity-b}
  2C_b=  -\omega\,\varphi(0)-\varphi'(0_+)+\vartheta'(0)
  =\omega\,\vartheta(\infty).
\end{equation}
Together, Eqns.~(\ref{eq:discontinuity-a})
and~(\ref{eq:discontinuity-b}) imply that
\begin{equation}
  \label{eq:CaCb}
  C_a=C_b={-\varphi'(0_+)/ 2}.
\end{equation}

Because of the sign function multiplying the first term in the l.h.s.,
Eq.~(\ref{eq:integrated-EL-A}) cannot be solved directly by a Fourrier
transformation.  It is, however, of the form solvable by the
Wiener-Hoph technique\cite{wiener-hopf-book}.  Following the standard
prescription, we introduce the functions $A_\pm(x)=A(x)\,\tf(\pm x)$,
so that, {\em e.g.}, $A(x)=A_+(x)+A_-(x)$,
$A(x)\sgn(x)=A_+(x)-A_-(x)$.  After this substitution we can
Fourrier-transform Eq.~(\ref{eq:integrated-EL-A}),
\begin{equation}
  \label{eq:fourrier-transformed-A}
  \omega [A_+-A_-]+ik\, Z(k)\, [A_++A_-] =2i\,C_a{\cal P}{1\over k},
\end{equation}
where ${\cal P}$ denotes the principal value, and the
Fourrier-transformed functions $A_\pm\equiv A_\pm(k)$ have no
singularities above and below the real axis respectively
(regularization at infinity ensures that they are also analytic
everywhere along the real axis).  The functions $A_\pm(x)$ are only
discontinuous in the origin, and the asymptotic form of their Fourrier
transformations at $|k|\to \infty$ is
\begin{equation}
  \label{eq:asymptotic-ug}
  A_\pm(k)=\pm{ i\over k}\,A_\pm(0_\pm)+{\cal O}(|k|^{-2})
  =\pm i\,{\varphi(0)\over2 k}+\ldots
\end{equation}

The independent functions in Eq.~(\ref{eq:fourrier-transformed-A}) can
be rearranged as follows,
\begin{eqnarray}
  A_+(k)&=&-{\cal R}(k)\, A_-(k)+{2C_a\over
   k\,Z-i\omega}\,{\cal P}{1\over k}, \label{eq:function-R}\\
  {\cal R}(k)&\equiv&
 {k\,Z+i\omega\over k\,Z-i\omega}={{\cal
     R}_-(k)\over{\cal R}_+(k)}\label{eq:fraction-decomposition}
\end{eqnarray}
where the function ${\cal R}(k)$ was separated into the ratio of the
function ${\cal R}_-(k)$ which has neither singularities nor zeros at
and below the real axis, and ${\cal R}_+(k)$, which has the same properties
at and above the real axis.  This separation is possible because the
function ${\cal R}(k)$ is analytic in a vicinity of the real axis
(which is correct for any $\omega$, assuming that the interaction
potential $V(x)$ is properly regularized at infinity). In the absence
of the long-distance interactions, $\chi=0$, the decomposition is
trivial, ${\cal R}_\pm^{0}=(k\pm i\omega)^{-1}$, where we assume
$\omega>0$.  At very large values of $k$ the long-distance part of the
potential should not matter.  Therefore, to ensure the regularity of
the decomposition~(\ref{eq:fraction-decomposition}) at $\chi>0$, we
can use the Cauchy formula
\begin{equation}
  \ln {r}_\pm(q)=\!\int_{-\infty}^\infty {dk\over2\pi i}\,
  {\ln {r}(k)\over q-k\pm i0},\;\;  {r}_\pm(q)\equiv{ {\cal
  R}_\pm(q)\over  {\cal R}^{0}_\pm(q)}
\label{eq:Cauchy-R}
\end{equation}
for the ratio $r(k)={\cal R}(k)/{\cal R}^{0}(k)$.  Since ${r}(k)\to1$
at large $k$, this expression implies that $r_\pm(k)\to1$ (and hence
that ${\cal R}_\pm\sim 1/k$) as $|k|\to\infty$.

Multiplying Eq.~(\ref{eq:function-R}) by ${\cal
  R}_+$, and separating the free term of the obtained expression into
a sum of functions analytic above and below the real axis
respectively, we obtain
\begin{equation}
  A_+(k)\,{\cal R}_+-2C_a\,h_+=-A_-(k)\,{\cal
    R}_-+2C_a\,h_-.
\label{eq:eqn-A-decomposed}
\end{equation}
Here the functions $h_\pm\equiv h_\pm(k)$, analytic in the upper
(lower) complex half-plane, are defined so that $h_+(k)+h_-(k)=h(k)$,
where
\begin{equation}
  \label{eq:function-h-decomposed}
  h(k)\equiv {{\cal R}_+(k)\over k\,Z-i\omega} \,{\cal P} {1\over k}
  ={{\cal R}_-(k)-{\cal R}_+(k)\over 2i\,\omega}\,{\cal P}{1\over k};
\end{equation}
these functions can be found  using the Cauchy formula
\begin{equation}
  h_\pm(q)=\mp {1\over 2\pi\, i}\,\int_{-\infty}^\infty\!
  {dk}\, {h(k)\over q-k\pm i0}.
  \label{eq:Cauchy-h}
\end{equation}
We assumed that ${\cal R}_\pm(k)$ are non-singular in the origin (and
elsewhere along the real axis), therefore, using the identity ${\cal
  R}_-(0)={\cal R}(0)\,{\cal R}_+(0)=-{\cal R}_+(0)$, we obtain
\begin{equation}
  \label{eq:h-plus-evaluated}
  h_\pm(k)=\pm i\,{{\cal R}_\pm(k)\over
    2\omega \,(k\pm i0)}. 
\end{equation}

By construction, the LHS of Eq.~(\ref{eq:eqn-A-decomposed}) has no
singularities at and above the real axis, while its RHS has no
singularities at and below the real axis.  Therefore, the whole
expression is analytic everywhere in the complex plane, and, as long
as it is uniformly limited at infinity, it can only be a constant.
Moreover, since both sides of Eq.~(\ref{eq:eqn-A-decomposed}) actually
{\em vanish\/} at infinity [as follows from
  Eq.~(\ref{eq:asymptotic-ug}) and the properties of the functions
  ${\cal R}_\pm$, $h_\pm$], this implies that the whole expression can
only be zero everywhere at the complex plane $k$.  We obtain
\begin{equation}
  \label{eq:solution-trivial}
  A_\pm(k)=2C_a\,{h_\pm(k)\over {\cal R}_\pm(k)}=\pm {i C_a\over
    \omega\, (k\pm i0)}, 
\end{equation}
and by matching with the asymptotic
expansion~(\ref{eq:asymptotic-ug}), we get
\begin{equation}
  C_a={\omega\,\varphi(0)\over2}, \quad A_\pm(k)=\pm {i
    \varphi(0)\over2\, (k\pm i0)}.\label{eq:A-found}
\end{equation}
Comparing to Eq.~(\ref{eq:CaCb}), we obtain
$$
\Delta\varphi'_0=2\varphi'(0_+)=-2\omega\varphi(0)
$$
and the contribution at the frequency $\omega>0$ to the effective
action~(\ref{eq:effective-action-evaluated}) becomes
$$
{\cal S}_{\rm q}(\omega)={T\over4\pi}
|\omega|\,|\varphi(0)|^2.  
$$
One can also obtain an identical contribution at $\omega<0$, so that
\begin{equation}
  \label{eq:K-half-pi-exact}
  {\cal K}_{\alpha=\pi/2}(\omega)= 1,
\end{equation}
as expected by the self-duality of the problem.

The analogue of Eq.~(\ref{eq:EL-A}) for the function $B(x)$ differs
only by the sign of $\omega$, which leads to a replacement ${\cal
  R}\to1/{\cal R}$, ${\cal R}_\pm\to 1/{\cal R}_\pm$.  Instead of
Eq.~(\ref{eq:eqn-A-decomposed}) we get
\begin{equation}
  \label{eq:eqn-B-decomposed}
  B_+(k)\,{\cal R}^{-1}_+-2C_b\,f_+=-B_-(k)\,{\cal
    R}^{-1}_-+2C_b\,f_-.
\end{equation}
By analogy with Eq.~(\ref{eq:h-plus-evaluated}), we obtain
\begin{equation}
  \label{eq:f-plus-evaluated}
  f_\pm(k)=\mp i\,{{\cal R}^{-1}_\pm(k)\over
    2\omega \,(k\pm i0)}. 
\end{equation}
By the same analyticity argument, both sides of
equation~(\ref{eq:eqn-B-decomposed}) are analytic everywhere in the
complex plane; at $|k|\to\infty$ they asymptotically approach a
constant value $i\varphi(0)$.  Therefore,
\begin{displaymath}
  B_\pm(k)= \pm{i \varphi(0)}\,\left[{\cal R}_\pm(k)-{1\over 2(k\pm
  i0)}\right] ,
\end{displaymath}
and, combining with Eq.~(\ref{eq:A-found}), we can use the
definition~(\ref{eq:a-b-defined}) to restore the original fields in
the extremum,
\begin{eqnarray}
  \label{eq:final-phi-half-pi}
  \varphi(x)&=&\varphi(0)\int {dk\over2\pi i}\left[ {\cal R}_-(k)-{\cal
        R}_+(k)\right] e^{-ikx},\\
  \vartheta(x)&=&\sgn(\omega\,x)\,[\varphi(0)-\varphi(x)],
  \label{eq:final-theta-half-pi}
\end{eqnarray}
where the $\sgn(\omega)$ in the second line is needed because the case
$\omega<0$ is equivalent to the interchange of $A$ and $B$, which
changes the sign of $\theta(x)$.

It is easy to verify that the obtained functions obey the boundary
conditions assumed when deriving
Eqns.~(\ref{eq:effective-action-evaluated}),
(\ref{eq:discontinuity-a}), (\ref{eq:discontinuity-b}).  This
self-consistency check ensures that the obtained expressions give us
the exact formal solution of the problem.

To understand the structure of this solution,
let us introduce the expansion
\begin{equation}
\chi V(x)=\sum_{l=1}^N {A_l\over a_l} \,e^{-a_l\,|x|}, \quad
\chi V(k)=\sum_{l=1}^N{2A_l\over
  k^2+a_l^2}, \label{eq:exponent-expansion}
\end{equation}
which, for sufficiently large $N$, gives an adequate regularized
representation of any non-pathological even function $V(x)$.  For
example, the Coulomb potential $ V(x)=1/|x|$ can be rewritten
as follows,
$$
{1\over |x|}=\lim_{a\to0}{a\over 1-\exp({-a |x|})}
=\lim_{a\to0}a\sum_{l=0}^\infty e^{-a l\,|x|}, 
$$
so that, given a finite $a$, any partial sum provides a
regu\-lar\-i\-zation of the form~(\ref{eq:exponent-expansion}) with
$a_l=a\,l$ and $A_l=\chi a^2l$.

We obtain
\begin{eqnarray*}
Z&=&1+{\chi\over2} V(k)=1+\sum_{l=1}^N {A_l\over k^2+a_l^2}, \\
k Z-i\omega&=&{P_{2N+1}(k) \prod_{l=1}^N (k^2+a_l^2)^{-1}},
\end{eqnarray*}
where the polynomial
$$
P_{2N+1}(k)= \prod_{s=1}^{2N+1}(k-i\kappa_s)
$$
has precisely $(2N+1)$ purely imaginary distinct roots $k_s\equiv
i\kappa_s\neq0$.  One can also show that for $\omega>0$ exactly $N$ of
the roots lie below the imaginary axis; we shall assume $\kappa_s<0$
for $1<s<N$.  The Cauchy integral~(\ref{eq:Cauchy-R}) is readily
evaluated, and we obtain
\begin{eqnarray}
  {\cal R}_+={(k-i\kappa_1)\ldots (k-i\kappa_N)\over 
    (k+i\kappa_{N+1})\ldots   (k+i\kappa_{2N+1})};
  \label{eq:exponent-expansion-R}
\end{eqnarray}
using the form similar to that in the first part of
Eq.~(\ref{eq:function-h-decomposed}), the extremum
solution~(\ref{eq:final-phi-half-pi}) can be explicitly rewritten as
\begin{equation}
  \varphi(x)=2|\omega|\,\varphi(0)\int {dk\over 2\pi}\,
  {(k^2+a_1^2)\ldots (k^2+a_N^2)\,\cos(kx)\over (k^2+\kappa^2_{N+1})\ldots
    (k^2+\kappa_{2N+1}^2)}.\label{eq:found-explicit-phi}
\end{equation}

\subsection{Expansion around the self-dual solution}
\label{sec:perturbation-theory}

To get an approximate expression for ${\cal K}(\alpha)$ in a vicinity
of $\alpha=\pi/2$, we expand $V_\pm(x,y)$ to first order in
$\cos\alpha$, and employ perturbation theory.  The solution of the
extremum equations at $\alpha_0=\pi/2$ is unique, and the lowest order
non-degenerate perturbation theory suffices.  This amounts to
evaluating the Euclidean action~(\ref{eq:symmetrized-action}) along
the non-perturbed solution~$\varphi(x)$, $\vartheta(x)$, 
\begin{eqnarray*}
  \delta{\cal S}_{\rm q}&\equiv&{T\over 4\pi}\sum_n|\omega_n|\,\delta {\cal
    K}_\alpha\,|\varphi(0)|^2\\
  &=&{T\over 4\pi}\sum_n
    {\chi\over2} \int dx \,dy  
    \left[\bar\varphi'_x\,\delta V_+\,\varphi'_y
    +\bar\vartheta'_x\,\delta V_-\,\vartheta'_y\right],
\end{eqnarray*}
where the integration is performed everywhere except the origin, and
the potentials
\begin{eqnarray*}
\delta V_+&=&-{ xy\,\cos\alpha\over \sqrt{x^2+y^2}}V'(\sqrt{x^2+y^2}),\\
 \delta V_-&=&-\delta V_+\,\sgn(xy). 
\end{eqnarray*}
were found by expanding Eqns.~(\ref{eq:v-plus}), (\ref{eq:v-minus}).

According to our solution~(\ref{eq:final-theta-half-pi}), the
functions $\varphi'(x)$, $-\vartheta'(x)\,\sgn(\omega\,x)$ are identical,
and the two terms give equal contributions, leading to
$$
\delta{\cal K}_\alpha=-{\chi\cos\alpha\over|\omega|\,|\varphi(0)|^2}
\int_{-\infty}^\infty\!  dx\,dy\,\bar\varphi'_x\,\varphi'_y\,{y}\,
\partial_x\,V(\sqrt{x^2+y^2}).
$$
For the Coulomb potential~(\ref{eq:coulomb}), this gives 
$$
\delta{\cal K}_\alpha={4\chi\cos\alpha\over|\omega|\,|\varphi(0)|^2}
\int_0^\infty\! dx\int_0^\infty\!dy\,
{x \,y\,\bar\varphi'_x\,\varphi'_y\over (x^2+y^2+a^2)^{3/2}}. 
$$
This integral converges at small distances even if we set $a\to0$; in
this scale-invariant limit the ``wavefunctions'' $\varphi(x)$ can
depend only on the dimensionless quantities $|\omega|x$ and $\chi$,
$\varphi(x)\equiv\varphi(0)\,\phi_\chi(|\omega|\,x)$.  Scaling out the
frequency leads to a {\em frequency-independent\/} correction,
\begin{eqnarray}
  \nonumber
  \delta{\cal K}_\alpha(\omega,\chi)&=&{\chi\,{\cal
      N}(\chi)\,\cos\alpha}+{\cal O}(\chi^2\cos^2\alpha),\quad
  \omega\,a\ll1,\\  
  {\cal N}(\chi)&\equiv& 4\!\int_0^\infty\!\!
  dx\int_0^\infty\!\!dy\, {x
    \,y\,\bar\phi'_\chi(x)\,\phi'_\chi(y)\over
    (x^2+y^2)^{3/2}}.  \label{eq:app-linear-expansion}
\end{eqnarray}
This result supports the numerical data, which indicates that
${\cal K}_\alpha(\omega)$ is {\em independent} of $\omega$ at small
enough frequencies.  This statement is true for all finite angles,
$|\cos\alpha|<1$, while ${\cal K}_{\alpha=0}(\omega)$ diverges
logarithmically according to Eq.~(\ref{eq:alpha0}).

The specific value of the correction depends on the coupling constant
$\chi$.  In the weak-coupling limit, $\chi\ll1$, the function
$\phi_{\chi\to0}(x)=\exp(-|x|)$, and the integration produces
$$
{\cal N}(\chi\to0)\approx 1.51.
$$
For finite $\chi>0$, and any given $N$ in the
expansion~(\ref{eq:exponent-expansion}), the explicit form of the
integrand in Eq.~(\ref{eq:app-linear-expansion}) can be found with the
help of Eq.~(\ref{eq:found-explicit-phi}), and the corresponding value
${\cal N}(\chi)$ can be evaluated numerically.

\end{document}